\documentclass[prb,twocolumn,showpacs,preprintnumbers,amsmath,amssymb]{revtex4-1}

\usepackage{graphicx}
\usepackage{dcolumn}
\usepackage{bm}

\begin{document}

\title{Superconductivity and Magnetism in YFe$_2$Ge$_2$}

\author{David J. Singh}

\affiliation{Materials Science and Technology Division,
Oak Ridge National Laboratory, Oak Ridge, Tennessee 37831-6056}

\date{\today}

\begin{abstract}
We report calculations of the electronic structure and magnetic
properties of YFe$_2$Ge$_2$ and discuss the results in terms of
the observed superconductivity near magnetism.
We find that YFe$_2$Ge$_2$ is a material near a magnetic
quantum critical point based on comparison of standard density
functional results that predict magnetism with experiment.
The band structure and Fermi surfaces are very three dimensional
and higher conductivity is predicted in the $c$-axis direction.
The magnetism is of Stoner type and is predominately from an
in-plane ferromagnetic tendency. The inter-layer coupling is weak
giving a perhaps 2D character to the magnetism, which is in
contrast to the conductivity and may be
important for suppressing the ordering tendency.
This is compatible with a triplet superconducting state mediated
by spin fluctuations.
\end{abstract}


\maketitle

\section{introduction}

The interplay of superconductivity and magnetism is a subject of 
long standing interest.
\cite{berk}
While the early interest was motivated by the observation that
nearness to magnetism is destructive to electron-phonon superconductivity,
as in e.g. elemental Pd, it was also realized that nearness to magnetism
could lead to new forms of unconventional
superconductivity with diverse order parameters.
\cite{berk,mathur}
Such cases can be particularly interesting, as 
exemplified by the Fe-pnictide and chalcogenide superconductors,
\cite{mazin-spm,kuroki-spm}
and perhaps the high $T_c$ cuprates.
\cite{scalapino,schrieffer}

The purpose of this paper is to examine the compound
YFe$_2$Ge$_2$, which was recently found to exhibit superconductivity
in close association with a magnetic state. \cite{zou}
YFe$_2$Ge$_2$ occurs in the ThCr$_2$Si$_2$ structure, is based
on Fe, and is near to magnetism.
This is similar to one of the main families
of Fe-based superconductors, specifically doped BaFe$_2$As$_2$ and
related compounds.
Unlike those compounds, the specific heat in YFe$_2$Ge$_2$ appears
to be more highly enhanced, the Wilson ratio is higher than two,
and the nearby magnetic order is of a different nature than that of
the Fe-based superconductors.
Several related $R$Fe$_2$Ge$_2$ ($R$=rare earth) compounds
show ordered antiferromagnetism.
\cite{pinto,fujiwara,avila,ferstl}
While this has been largely discussed as rare-earth magnetism,
the fact that it also occurs in the Lu compound, which has no
rare-earth moment indicates that it actually involves the Fe
as well. \cite{fujiwara}
Ishida and co-workers, \cite{ishida} anticipated a nearness to
magnetism early on using density of states arguments, which
we confirm in detail here.

\section{approach}

Here, we report calculations of the electronic structure and magnetic
behavior of YFe$_2$Ge$_2$ and discuss these in relation to the
superconductivity.
Our calculations are done within density functional theory, similar
to the recently reported work of Subedi. \cite{subedi}
The present calculations were done using the general potential
linearized augmented planewave code, with local orbitals,
\cite{singh-book} as implemented in the WIEN2k code. \cite{wien2k}
We used LAPW sphere radii of 2.5 bohr, 2.2 bohr and 2.2 bohr,
for Y, Fe and Ge, respectively.
Semicore states (Y 4$s$, Y 4$p$, Fe 3$p$ and Ge 3$d$) were included
with the valence states. We used the LAPW plus local orbitals basis set
and a well converged planewave cutoff set at $RK_{max}$=9
(here $K_{max}$ is the planewave cutoff and $R$ is the smallest
LAPW sphere radius, i.e. 2.2 bohr).
The Brillouin zone samplings were done using uniform grids and convergence
with respect to these grids was tested.
The exchange correlation functional of Perdew, Burke and Ernzerhof
(PBE) was employed. \cite{pbe}

For the structure, we used the experimental lattice parameters of
Zou and co-workers, i.e. $a$=3.9617 \AA, and $c$=10.421 \AA;
Y at (0,0,0), Fe at (0,1/2,1/4) and (1/2,0,1/4) and
Ge at (0,0,$z$) and (0,0,1-$z$) plus the additional
body centered positions ($x$,$y$,$z$) $\rightarrow$
($x$+1/2,$y$+1/2,$z$+1/2).
These lattice parameters are similar to those in the earlier 
experimental report of Venturini and Malaman. \cite{venturini}
The internal coordinate corresponding to the Ge position in 
the unit cell was determined by total energy minimization.
We did this in two ways -- non-spin-polarized, corresponding to the
experimental paramagnetic state, and spin-polarized in the $A$-type
antiferromagnetic ground state configuration of the related compounds,
$R$Fe$_2$Ge$_2$, which is also the ground state found in the present
calculations (see below).

The magnetic calculation yielded $z$=0.373, for an Fe-Ge 
neighbor distance of 2.358 \AA,
while the non-spin-polarized calculation yielded $z$=0.370,
for an Fe-Ge neighbor distance of 2.343 \AA.
This difference indicates a non-negligible magnetoelastic coupling,
but still much smaller than the giant effects found in similar
calculations for the Fe-based superconductors.
\cite{mazin-mag}
The results shown below are for the magnetic value.

\section{electronic structure}

\begin{figure}
\includegraphics[width=\columnwidth,angle=0]{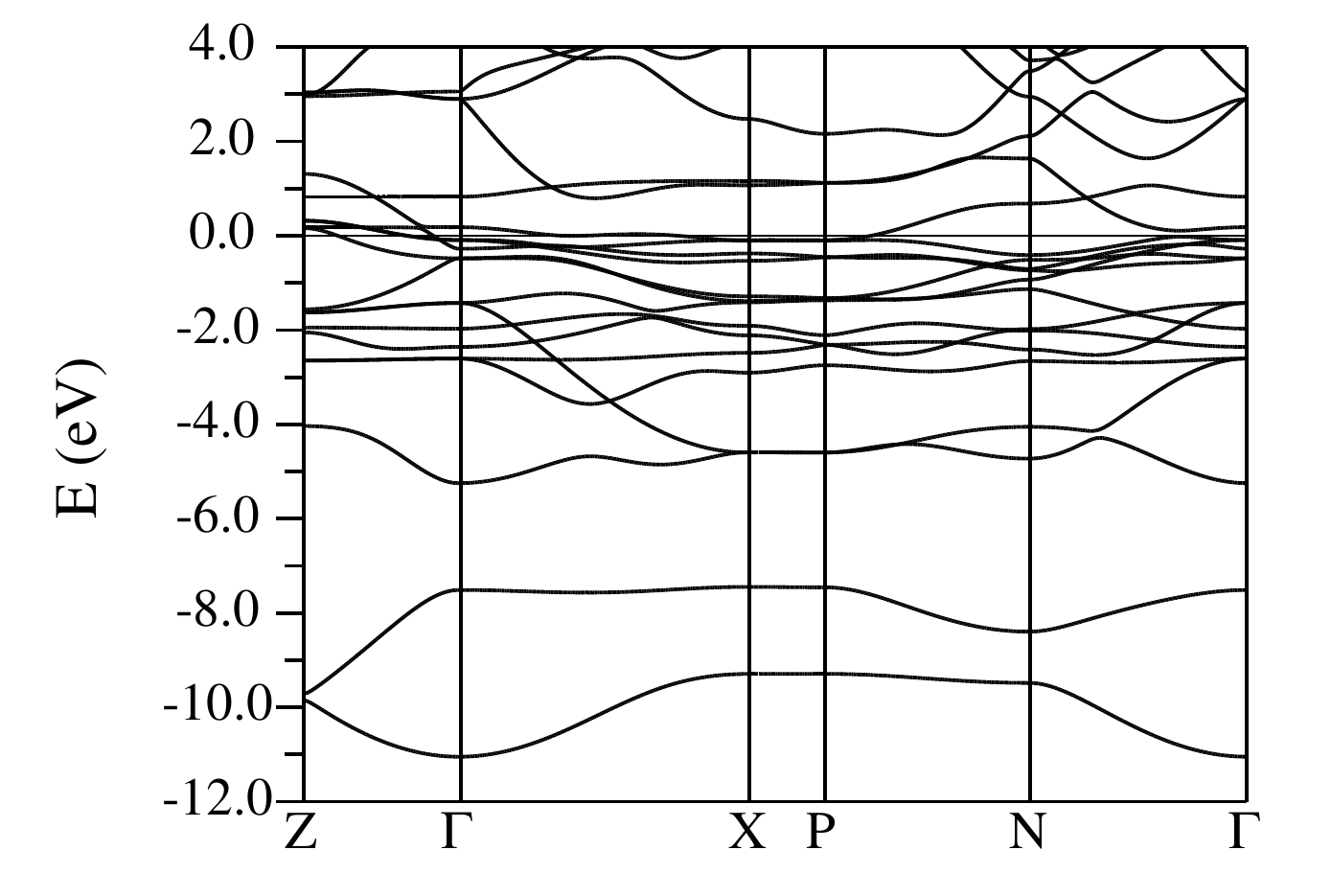}
\caption{Band structure of YFe$_2$Ge$_2$ with the Fermi level
at 0 eV. The two deep bands centered at $\sim$ -9 eV are from the
Ge 4$s$ state, while the higher valence bands arise primarily
from hybridized Ge $4p$ - Fe $3d$ states.}
\label{bands}
\end{figure}

YFe$_2$Ge$_2$ may be notionally related to the iron-superconductors
via electron count. Specifically, in comparison to SrFe$_2$As$_2$,
the replacement of As by Ge and Sr by Y would lead to a deficit of
one electron per cell. Then YFe$_2$Ge$_2$ would be notionally
like SrFe$_2$As$_2$ doped by 0.5 holes per Fe. Interestingly, KFe$_2$As$_2$,
which has the same electron count, is a low temperature superconductor
near magnetism like YFe$_2$Ge$_2$.
\cite{chen}

\begin{figure}
\includegraphics[width=\columnwidth,angle=0]{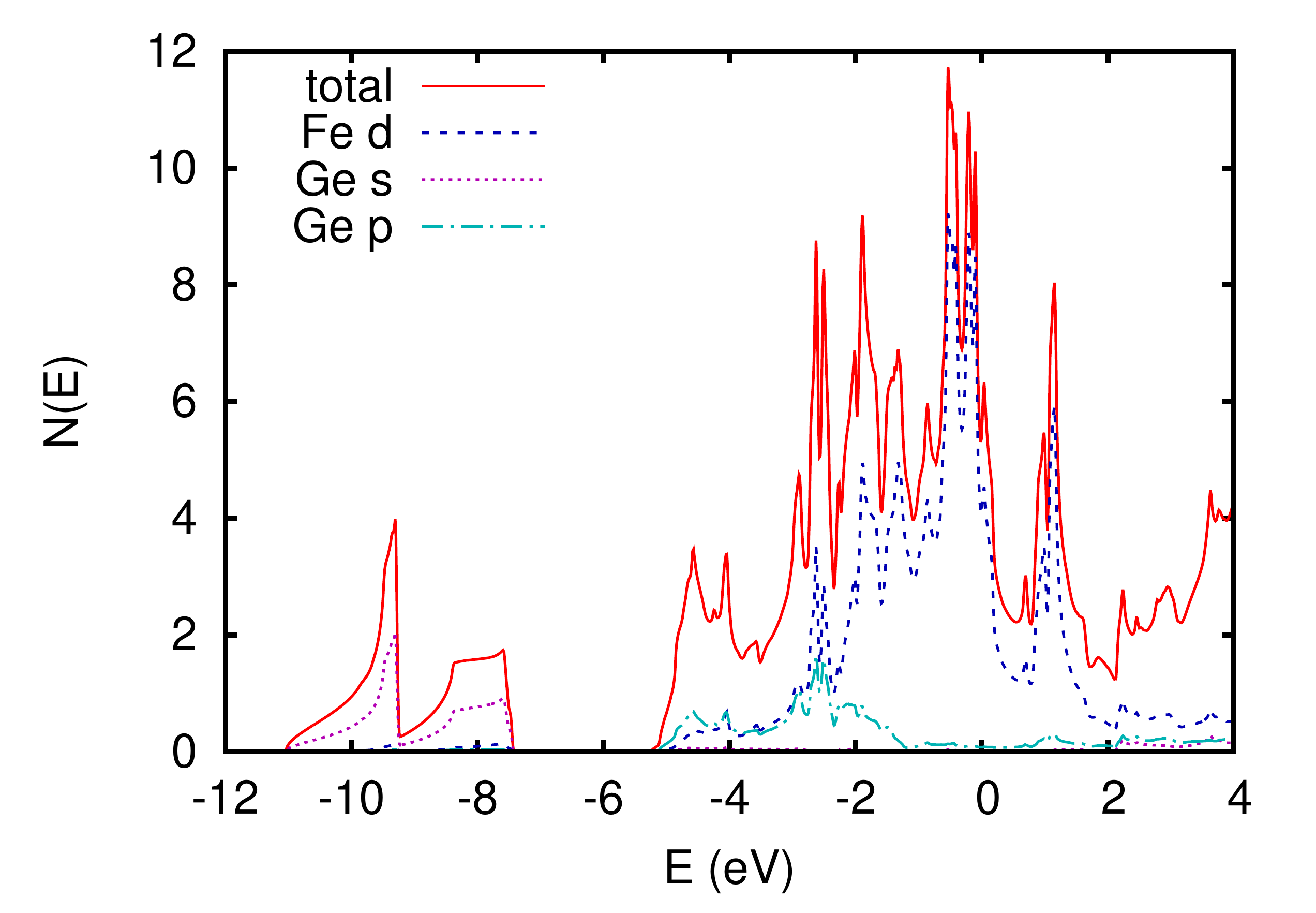}
\caption{(color online) Density of states and projections onto LAPW
spheres.}
\label{dos}
\end{figure}

\begin{figure}
\includegraphics[width=\columnwidth,angle=0]{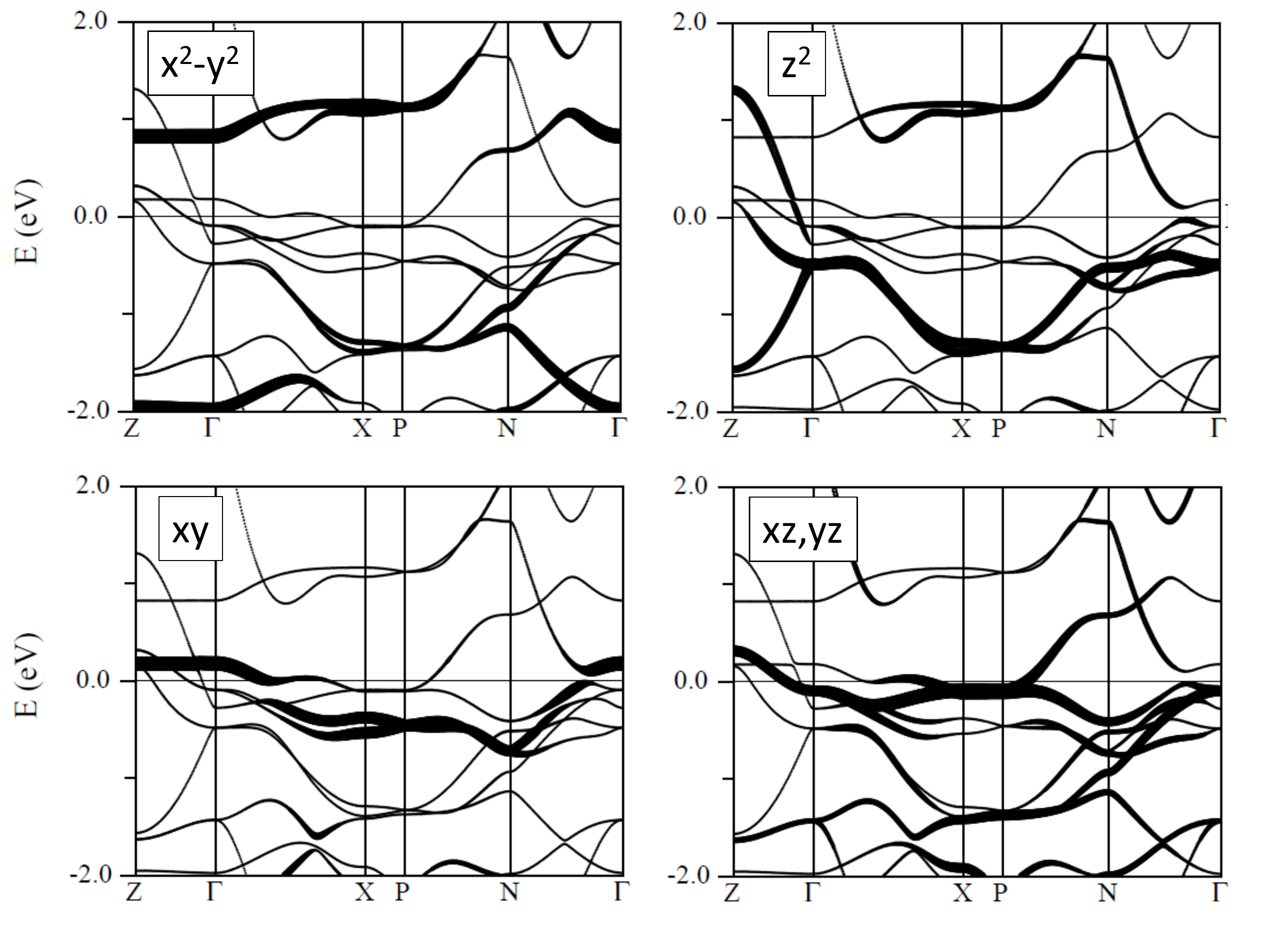}
\caption{Band structure of YFe$_2$Ge$_2$ around the Fermi
energy (0 eV, light dotted line), showing the different orbital characters via
so-called fat bands plots, in which the size of the plotting
points are proportional to the given orbital characters plus
a small constant.}
\label{bands-proj}
\end{figure}

The band structure and corresponding electronic density of states
are shown in Fig. \ref{bands} and Fig. \ref{dos}.
The valence band electronic structure derives from hybridized Ge 4$p$ - Fe 3$d$
states, similar to the Fe-pnictide superconductors, and there
is dominant Fe 3$d$ character from $\sim$ -3 eV to $\sim$ 2 eV
relative to the Fermi energy, $E_F$. Similar to hole doped Fe-pnictides,
there is a dip in the density of states above $E_F$ and there is a 
high $N(E_F)$. The calculated value is $N(E_F)$=5.26 eV$^{-1}$ on a per
formula unit basis. This corresponds to a bare
Sommerfeld specific heat coefficient $\gamma_{\rm bare}$ =
12.4 mJ mol$^{-1}$K$^{-2}$.
This is about eight times smaller than the experimental value of 
$\gamma$$\sim$ 100 mJ mol$^{-1}$K$^{-2}$.
As discussed by Zou and co-workers, such a high value could be
due to nearness to a magnetic quantum critical point.
Interestingly, KFe$_2$As$_2$ displays a similarly high value.
\cite{sato-gamma}
The Fe $d$ contribution to $N(E_F)$, as measured by projection onto
the Fe LAPW spheres is 3.99 eV$^{-1}$ per formula unit, i.e. $\sim$
2 eV$^{-1}$ per atom, which places the material near Stoner itinerant
magnetism.
As shown in Fig. \ref{bands-proj},
this Fe $3d$ contribution to $N(E_F)$ comes from multiple orbitals
again similar to the Fe-pnictides.
\cite{singh-fesc}

The electronic structure and properties are, however,
otherwise very different from those of KFe$_2$As$_2$.
\cite{singh-kco2as2,sato}
This implies that the physics may also be very different
from the Fe-pnictide superconductors.

First of all the electronic structure is very three dimensional and
has strong dispersion near $E_F$ in the direction perpendicular to the
Fe sheets.
As noted by Subedi, \cite{subedi} the compound has significant
Ge-Ge bonding. This can explain the high band dispersion perpendicular
to the layers (note that the Shannon ionic
radius of trivalent Y is 1.04 \AA, i.e.
the atom in the layer between the Ge is much smaller than those of
the corresponding atoms in the Fe-pnictides,
Sr (1.32 \AA), Ba (1.49 \AA) or K (1.52 \AA)).
The calculated plasma frequencies are
$\hbar\Omega_{p,xx} = \hbar\Omega_{p,yy}$ = 2.83 eV and
$\hbar\Omega_{p,zz}$ = 4.41 eV.
Thus the high conductivity direction is predicted to be perpendicular
to the planes, with a sizable anisotropy
$\sigma_{zz}/\sigma_{xx} \sim$ 2.4 in the constant scattering
time approximation.

\begin{figure}
\includegraphics[width=\columnwidth,angle=0]{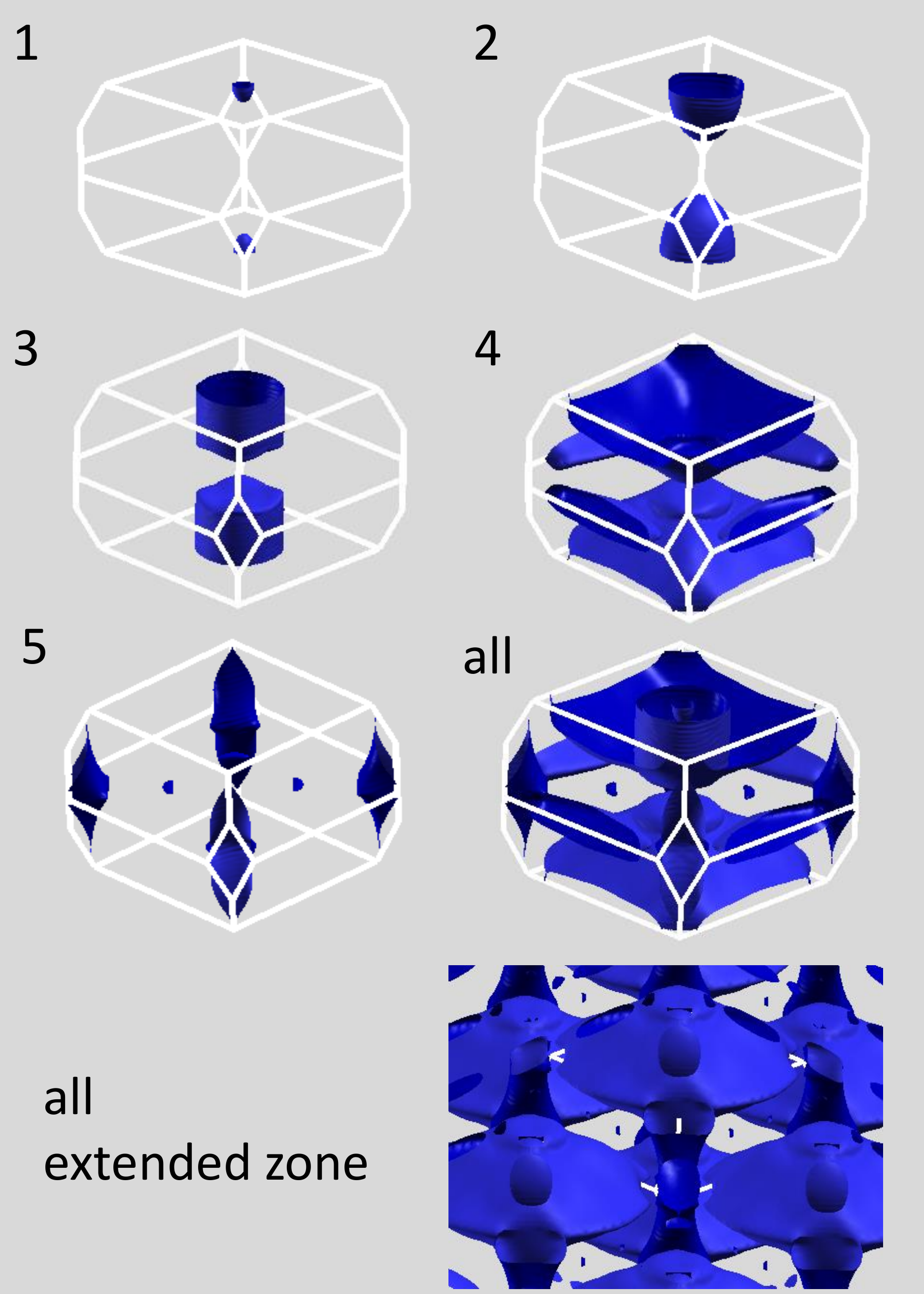}
\caption{(color online) Calculated Fermi surfaces of 
YFe$_2$Ge$_2$, showing the five sheets (1-5). The bottom
shows the Fermi surface in an extended zone scheme.}
\label{fermi}
\end{figure}

The Fermi surfaces are shown in Fig. \ref{fermi}.
While there are minor differences from those reported by Subedi, \cite{subedi}
the important aspects are the same.
There are three main sheets of Fermi surface, ``3", ``4",
and ``5" in Fig. \ref{fermi},
plus several small sheets. The main sheets are a truncated hole cylinder
centered at the $Z$ point, a large disk section that touches the edges
of the zone, and electron cylinders at the zone corners.
The small surfaces are tiny ellipsoid around $Z$, a
hole ellipsoid around $Z$,
and tiny pieces from band extrema
near $E_F$ along $\Gamma$-$X$.

The tiny ellipsoid
only $\sim$0.002 holes per cell, and is therefore negligible.
The second ellipsoid contains 0.08 holes, and the truncated hole cylinder
contains 0.15 holes. The electron cylinders at the zone corners
have 0.09 electrons and the remainder is the disk. This is the
dominant Fermi surface
and comes from a near half-filled, but electron doped
(filling 1.14 e) band. As seen in the band structure, this comes from
a light band of hybridized Fe $d_{z^2}$ - Ge $p$ character.
The truncated cylinder and the outer ellipsoid have $d_{xz}$,$d_{yz}$
character, as do the cylinders at the zone corners.
character (here we use a coordinate system where $z$ is perpendicular
to the layers and $x$ and $y$ point to the neighboring Fe atoms).

The flat parts of the two larger Fermi surfaces, the truncated cylinder
and the disk, are at $k_z$ of
0.34 and 0.17 of the distance from $\Gamma$ to $Z$, i.e. not at the nesting
distance of 0.5 for alternating planes along $k_z$. This contradicts
the conjecture that the antiferromagnetic ordering tendency of the
$R$Fe$_2$Ge$_2$ compounds is due to a spin-density wave associated
with Fermi surface nesting.
In any case, it is clear that the electronic structure and therefore
properties of YFe$_2$Ge$_2$ are dominated by a main disk shaped
Fermi surface, which is near half filling but electron doped and
is centered around the $Z$-point.

\section{magnetism}

As mentioned, the high $N(E_F)$ by itself suggests nearness to
itinerant magnetism. We do in fact find magnetism in our PBE
density functional theory calculations.
This magnetism is in contrast to experimental situation, where YFe$_2$Ge$_2$
is near magnetism, but remains a non-magnetically ordered
renormalized paramagnet down to low temperatures.
This situation is qualitatively similar
to what was found in the Fe-pnictide superconductors, where it
indicates nearness to a magnetic quantum critical point.
\cite{mazin-mag,bondino}
In general
this type of overestimate of magnetic tendencies within density
functional calculations is unusual. It typically arises when
spin fluctuations associated with a nearby quantum critical point are strong
enough to renormalize the mean-field like magnetic state predicted
by standard approximate density functional calculations.
\cite{moriya-book,mazin-zrzn2}

\begin{figure}
\includegraphics[width=\columnwidth,angle=0]{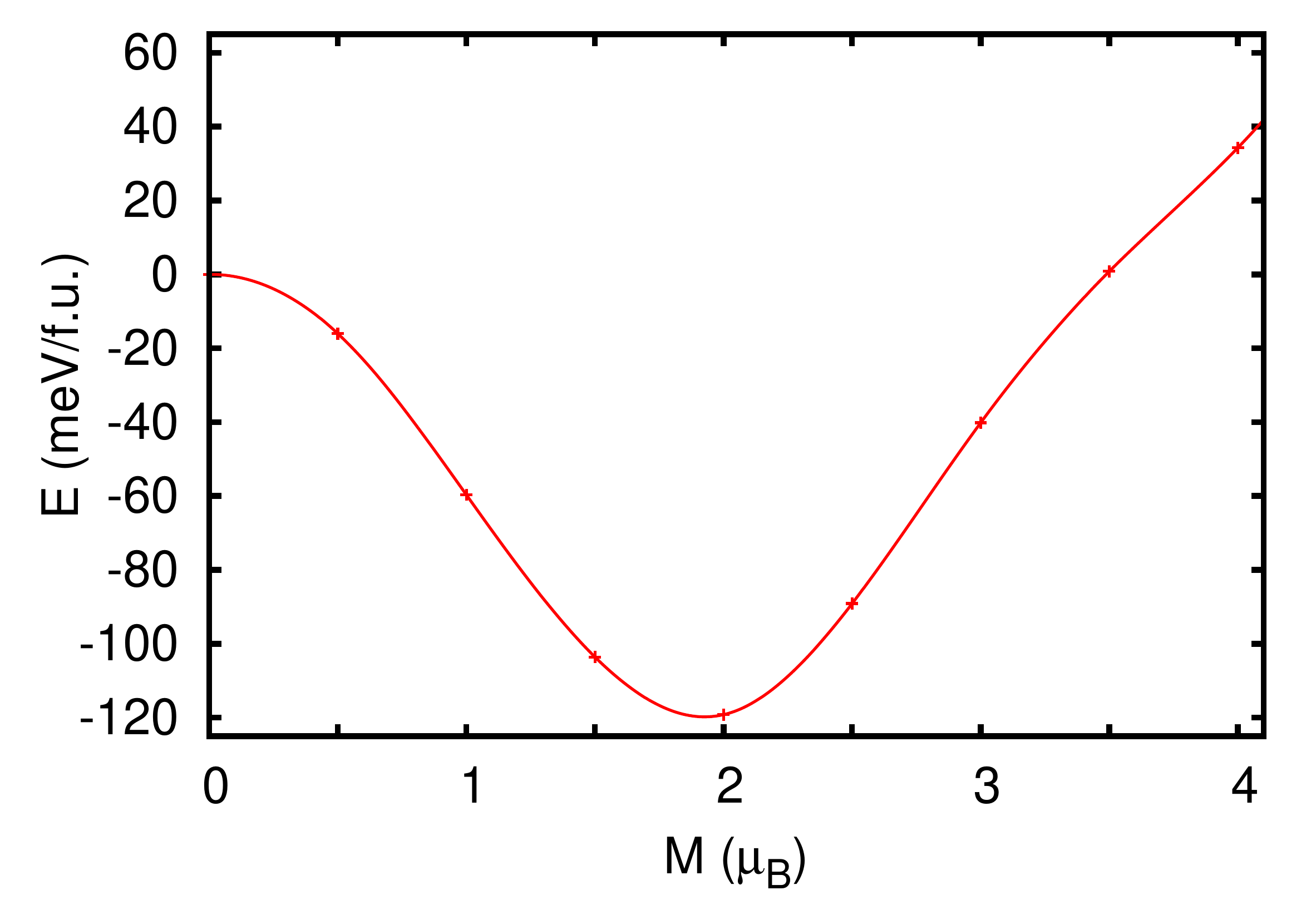}
\caption{(color online) Fixed spin moment total energy as a function
of constrained spin magnetization for YFe$_2$Ge$_2$ on a per
formula unit basis. The symbols are the calculated
points and the curve is a spline interpolation. Note the instability of the
non-spin-polarized state against ferromagnetism.}
\label{fsm}
\end{figure}

The results of fixed spin moment calculations are shown in Fig. \ref{fsm}.
As seen there is a sizable ferromagnetic instability, which can be
understood as a Stoner instability. This amounts to a spin magnetization
of 1.93 $\mu_B$ per cell, consisting of a polarization inside each Fe LAPW
sphere of 1.03 $\mu_B$ accompanied by a small back-polarization of the Ge.
The magnetic energy is large, $\sim$ 120 meV per formula unit.
The result is a spin-polarization of the bands, and a lowering of the
overall $N(E_F)$ to 3.9 eV$^{-1}$
(0.97 eV$^{-1}$ and 2.97 eV$^{-1}$,
for majority and minority spin, respectively).

We also did calculations for other magnetic orders. These were
an A-type order, where ferromagnetic Fe planes are stacked
antiferromagnetically along the $c$-axis direction,
a C-type order, where the Fe are arranged in a checkerboard fashion
in-plane, and stacked in ferromagnetic chains along $c$ and
a G-type order, with nearest Fe antiferromagnetic both in-plane and
along $c$. The resulting energies, moments and $N(E_F)$ are summarized
in Table \ref{tab-mag}.
As may be seen, the lowest energy state is the A-type, which corresponds
to the experimental antiferromagnetic state for LuFe$_2$Ge$_2$.

An examination of the energies of the different calculated states
shows an itinerant aspect.
In particular, the energy differences between
the different ordered states are of the same magnitude as the
energy differences between
the ordered states and the non-spin polarized case, although all the magnetic
configurations that we tried do form and have at least slightly
lower energy than the non-spin-polarized case.
The second apparent feature is that the A-type and ferromagnetic
states are close in energy, while the other two states are also
close to each other in energy, but are much above the ferromagnetic
state. 

Thus the primary driver for magnetism is an in-plane
ferromagnetic tendency associated with the high $N(E_F)$
of the non-spin-polarized state.
The interlayer interactions are apparently much weaker as evidenced 
by the similar energies of the ferromagnetic and A-type ordered states
and of the C-type and G-type states. The primary magnetic interactions
are in-plane reflecting a layered crystal structure, although the
conductivity is predicted to be highest out of plane.
Thus YFe$_2$Ge$_2$ is a very three dimensional metal that nonetheless
is predicted to have a more two dimensional magnetic behavior.

It is
notable that experimental measurements for the
closely related LuFe$_2$Ge$_2$
compound, which as mentioned
is antiferromagnetic,  shows Fe moments that lie in the
basal plane. \cite{fujiwara}
This situation
with in-plane moments, ferromagnetic interactions in-plane and weak
out-of-plane interactions suggests a scenario in which the ordering
temperature may be reduced by dimensional effects
(specifically, with in plane anisotropy on a square lattice
and weak coupling between the planes,
one may have a depression of the ordering temperature from that which
would be anticipated based on the strength of the
in-plane exchange interactions).
In this regard, Ferstl and co-workers, who
did specific heat and susceptibility measurements
for the related compounds LuFe$_2$Ge$_2$ and
YbFe$_2$Ge$_2$ report evidence for large fluctuating
Fe moments high above the ordering temperature. \cite{ferstl}
As seen, $N(E_F)$ is substantially reduced from the non-spin-polarized value by
in-plane ferromagnetism
(i.e. ferromagnetic and A-type antiferromagnetic order) but not by orders
that involve in-plane antiferromagnetism, consistent with the
Stoner mechanism.
The primarily magnetic tendency that we find is towards in-plane
Stoner magnetism, and this, and not a spin-density-wave, is
the reason for the moment formation.

In any case, this picture of the magnetism has certain consequences.
First of all, one may expect
metamagnetic transitions in the $R$Fe$_2$Ge$_2$
compounds under high field including
the non-magnetic compound YFe$_2$Ge$_2$.
These may be accompanied by a sizable magnetoresistance,
which should be negative in the range where ferromagnetic
order becomes imposed by the field.
These have been observed in
some of the compounds. \cite{avila}
Secondly, one expects a susceptibility, $\chi$({\bf q}) that shows
weak $k_z$ dependence and a stronger in-plane dependence peaked near the
2D zone center (and highest at $Z$). As noted by Subedi, there is also a 
nesting of the small cylinder sheets that can modify this by the addition
of a nesting related peak, which would couple the zone center to the zone
corner, i.e. an in-plane pattern similar to the Fe-pnictide superconductors,
but this would not couple to the main disk shaped Fermi surface.

\begin{table}
\caption{Magnetic energies, $E$, moments
$m$ inside the Fe LAPW spheres, radius, 2.2 bohr,
and $N(E_F)$ on a per formula unit basis
for different ordering patterns (see text). The
energy zero is taken for the non-spin-polarized case, denoted
``NSP".}
\begin{tabular}{lccc}
\hline
Order & $m$ ($\mu_B$/Fe) & $E$ (eV/f.u.) & $N(E_F)$ (eV$^{-1}$) \\
\hline
NSP & 0.00 & 0.000  & 5.3 \\
F   & 1.03 & -0.120 & 3.9 \\
A   & 1.04 & -0.137 & 3.6 \\
C   & 1.26 & -0.026 & 7.0 \\
G   & 0.98 & -0.004 & 6.4 \\
\hline
\end{tabular}
\label{tab-mag}
\end{table}

\section{discussion}

The magnetism has consequences for superconductivity.
Almost all superconductors are conventional $s$-wave superconductors
mediated by electron-phonon interactions, and this may also be the
case here.
However, there are two features that suggest consideration of
other possibilities. First of all, the specific heat $\gamma$
is very high. This suggests a short coherence length in which case
Coulomb avoidance should work against a conventional $s$-wave
state. Secondly, the nearness to a magnetic quantum critical point with
a ferromagnetic character suggests the presence of ferromagnetic spin
fluctuations, which are pair breaking for a singlet superconductor.
This means that if YFe$_2$Ge$_2$ is a conventional electron phonon
superconductor, it is one in which the superconductivity is heavily affected
by magnetism and which
would have a considerably higher critical temperature without magnetism.

An alternate scenario to the electron-phonon picture is spin-fluctuation
mediated superconductivity. This depends on the interplay between the
{\bf q}-dependent spin fluctuations, as characterized by the real part
of $\chi({\bf q})$, and the Fermi surface. Spin-fluctuations provide a
repulsive interaction for singlet superconductivity and an attractive
interaction for triplet superconductivity. \cite{berk}
The resulting superconducting state is then due to matching
of the {\bf q} dependence of $\chi({\bf q})$ with the Fermi surface.

In the present case, the major Fermi surface is the disk around the
$Z$-point. In a singlet channel one could imagine that spin-fluctuations
associated with the antiferromagnetic order (i.e. the antiferromagnetic
interaction along the $c$-axis) could couple the two faces of the 
disk. In that case, since one has a repulsive interaction, a state
in which the two faces have opposite order parameter would be favored.
However, because of the symmetry of the $Z$ point this would lead
to odd parity, i.e. not a singlet, while in a triplet channel the 
argument works in reverse -- the antiferromagnetic tendency would
favor having the same sign order parameter on the two faces, which would
then have even parity and not be a triplet. Therefore we conclude that
the antiferromagnetic interaction along $c$ is not effective in providing
pairing. In any case, as shown by the energies in Table \ref{tab-mag},
this interaction is not particularly strong.

Ferromagnetic spin
fluctuations are pair breaking for singlet superconductivity,
since they imply a repulsive interaction at low {\bf q} for
a singlet. In the triplet
case they are attractive at low {\bf q} and superconductivity can
arise if the susceptibility falls off on the scale of the Fermi surface
size. In the present case, the disk Fermi surface is large, and so it
can be anticipated that a triplet state in which the order parameter
changes sign going around the Fermi surface will be stabilized.
This could be of the $p$-wave type
proposed for Sr$_2$RuO$_4$,
\cite{rice,mackenzie}
which in this case would be a vector order parameter rotating as
one goes around periphery of the disk,
or perhaps a more complicated triplet state.

\section{summary and conclusions}

Thus we find that YFe$_2$Ge$_2$ is a material near a magnetic
quantum critical point based on comparison of standard density
functional results that predict magnetism with experiment.
The band structure and Fermi surfaces are very three dimensional
and higher conductivity is predicted in the $c$-axis direction.
The magnetism is of Stoner type and is predominately from an
in-plane ferromagnetic tendency. The inter-layer coupling is weak
giving a perhaps 2D character to the magnetism, which is in
contrast to the conductivity and may be
important for suppressing the ordering tendency.
Based on matching
of the Fermi surface with the magnetic tendency, it seems most
likely that YFe$_2$Ge$_2$ is either an electron-phonon superconductor,
in which case superconductivity must be strongly suppressed by the
magnetic tendency, or a triplet superconductor mediated by
the near ferromagnetic spin fluctuations acting on the large
Fermi surface. Considering the heavy mass implied by specific heat
measurements, the strong mass renormalization,
the experimental $R_W > 2$, and the very close
proximity to ferromagnetism, it seems that the triplet scenario may
be realized. Experiments that can distinguish these cases are
(1) correlating the critical temperature $T_c$ with the mean
free path when limited by paramagnetic impurities, i.e. inverse
correlations between resistivity and $T_c$;
(2) Specific heat measurements, since with the complex three dimensional
Fermi surface of YFe$_2$Ge$_2$ a triplet state may not be fully gapped;
(3) Spin susceptibility below $T_c$, e.g. using Knight shift;
and (4) searches for time reversal symmetry breaking.
\cite{luke,jia}

\acknowledgments

This work was supported by the U.S. Department of Energy,
Basic Energy Sciences, Materials Sciences and Engineering Division.

%\bibliography{YFe2Ge2}

\end{document}